\documentstyle[amssymb,aps,preprint]{revtex}

\begin{document}
\draft
\author{Daole Yin\cite{atr}, Zhi Qi, Hengyi Xu and Furen Wang}
\address{Department of Physics, Peking University, Beijing 100871, P. R. China}
\author{Lan Yin}
\address{Department of Physics, University of Washington, Box 351560 Seattle,
WA98195-1560}
\title{Resistive Transition Equation of Mixed State: A Microscopic Analysis}
\maketitle

\begin{abstract}
We present a microscopic derivation of the resistive transition equation for
mixed state of superconductors. This form fits the experimental data of $%
MgB_2$ with parameters in agreement with the prediction of BCS
superconductivity. It also fits the experimental data of high quality
untwinned YBCO single crystal but with parameters somewhat different from
the BCS prediction. A discussion in connection with the problem of cuprate
superconductivity is given.
\end{abstract}


\bigskip

I. INTRODUCTION

The discovery of high-temperature superconductors (HTS) revived the interest
in the mixed state physics. One of the most interesting issues is the
resistive transition behavior of a superconductor. The heuristic pioneering
work of Tihkham \cite{1} showed that by using a proper activation energy for
flux flow the experimentally measured early field-depending broadening of
HTS \cite{2} can be accounted for nicely by the model of thermally activated
flux flow model (TAFF) \cite{3}. However, subsequent measurements with
higher sensitivity showed that the fit was quantitative only down to one
tenth of the normal resistance $R\sim 0.1R_n$. A number of authors suggested
that the much more rapid drop of resistance than the exponential predicted
by the TAFF model might be due to a freezing / melting transition between a
vortex liquid and a vortex solid of some sort \cite{4}. Koch et al. \cite{5}
and subsequently Gammel et al. \cite{6} carefully tested experimentally the
predictions of the `vortex-glass model' proposed by M. P. A. Fisher \cite{7}
by measuring the $I\sim V$ curves and resistance of YBCO epitaxial thin
films and single crystals with high sensitivity. They found that for each
filed $B$ the isothermal $I\sim V$ curve shows a power-law behavior at a
uniquely defined temperature $T_g\left( B\right) $ as

\begin{equation}
V\propto I^n\text{.}  \eqnum{1a}
\end{equation}
At $T>T_g\left( B\right) $, the curvature on a $\ln V$ vs $\ln I$ plot is
positive corresponding to the form of TAFF models prediction

\begin{equation}
V\propto \sinh \left( \frac I{I_0}\right) \text{.}  \eqnum{1b}
\end{equation}
In contrast, at $T<T_g\left( B\right) $ one finds a negative curvature with
the characteristic predicted by vortex-glass model \cite{7}

\begin{equation}
V\propto \exp \left[ -\left( \frac{J_T}J\right) ^\mu \right] .  \eqnum{1c}
\end{equation}
Since $T_g\left( B\right) $ is defined as the dividing point between
temperatures for which the linear resistance (i.e., $R$ in the limit of $%
I\rightarrow 0$) is zero, as indicated by Eq. (1c), and that in which it is
not zero, as Eq. (1b), this transition temperature $T_g\left( B\right) $ is
operationally very much the same as the melting temperature of vortex solid $%
T_m\left( B\right) $ measured in (nearly) ideal crystals. Kwok et al.
observed a sharp `kink' or `knee' in the magnetic-field-broadened resistive
transition at $R/R_n\lesssim 0.12$ in a high-quality untwinned YBCO single
crystal. This behavior obeys the angular dependence expected from the
Lindemann criterion of vortex lattice melting \cite{8}. The resistive
transition broadening as well as the `irreversibility line' were also
observed in the recently discovered $MgB_2$ with $T_c\sim 40K$ \cite{9},
though it shows typical behavior of a phonon-mediated BCS superconductor
through the $B$ isotope effect \cite{10}.

Recently, it is found that the resistive transition of both HTS and $MgB_2$
can be well described with a equation of the normalized form

\begin{equation}
\frac R{R_n}=\exp \left[ -\sum_{i=1}^2\gamma _i\left( 1+y_i-x_i\right)
^{p_i}\theta \left( T_i-T\right) \right] \text{,}  \eqnum{2}
\end{equation}
with $x_i$, $y_i$, $\gamma _i$ and $T_i$ the normalized current, voltage,
symmetry-breaking factor and critical temperature respectively defined as

\begin{eqnarray}
x_1 &\equiv &\frac I{I_d\left( T\text{, }B\right) }\text{, } 
\begin{array}{cccc}
&  &  & 
\begin{array}{lll}
&  & 
\begin{array}{l}
\end{array}
\end{array}
\end{array}
x_2\equiv \frac I{I_{c0}\left( T\text{, }B\right) }  \nonumber \\
y_1 &\equiv &\frac R{R_n}\left[ \frac I{I_d\left( T\text{, }B\right) }%
\right] \text{,} 
\begin{array}{cc}
& 
\end{array}
\text{ }y_2\equiv \frac R{R_f}\left[ \frac I{I_{c0}\left( T\text{, }B\right) 
}\right]  \nonumber \\
\gamma _1 &\equiv &\ln \frac{R_n\left( T\text{, }B\right) }{R_f\left( T\text{%
, }B\right) }\text{, } 
\begin{array}{ccc}
&  & 
\begin{array}{lll}
&  & 
\end{array}
\end{array}
\gamma _2\equiv \frac{U_c\left( T\text{, }B\right) }{kT}  \nonumber \\
T_1 &\equiv &T_c\left( B\right) \text{, } 
\begin{array}{ccccc}
&  &  &  & 
\begin{array}{llllll}
&  &  &  &  & 
\end{array}
\end{array}
T_2\equiv T_m\left( B\right)  \eqnum{3}
\end{eqnarray}
where $I_d$ is the depairing current and $I_{c0\text{ }}$is the critical
current of vortex solid for overcoming the activation energy barrier $%
U_c\left( T\text{, }B\right) $. $\theta \left( x\right) $ is the Heaviside
function and $p_i$ are exponents \cite{11}.

In present work we try to show the connection of the resistive transition
equation (2) with the basic Ginzburg-Landau (G-L) theory \cite{12}which can
be derived from the microscopic theory of inhomogeneous superconductor as
shown by Gorkov \cite{13}.

In sections II and III we shall derive the two terms in the right hand side
braket of Eq. (2), i.e., $i=1$ and $2$, in connection with the
normal-superconducting state (N-S) transition and flux pinning respectively.
The derived equation will be compared with the experimental data of $MgB_2$
and YBCO single crystal in section IV. A discussion about possible
implications is given in section V.

II. TRANSITION NEAR $B_{c2}\left( T\right) $

The basic frame for describing the physics of superconducting mixed state is
the Ginzburg-Landau (GL) free energy density $f$ in the form \cite{12}

\begin{equation}
f=f_{n0}+\alpha \left| \Psi \right| ^2+\frac \beta 2\left| \Psi \right| ^4+%
\frac 1{2m^{*}}\left| \left( \frac \hbar i\nabla -\frac{e^{*}}c{\bf A}%
\right) \Psi \right| ^2+\frac{h^2}{8\pi }\text{,}  \eqnum{4}
\end{equation}
where $f_{n0}$ is the free energy density of normal state, $\Psi \left(
r\right) $ is the complex order parameter, ${\bf A}$ is the vector
potential. Gorkov found that the GL theory based on Eq. (4) is derivable as
a rigorous limiting case of the BCS microscopic theory with $\Psi \left(
r\right) $ proportional to the local value of the gap parameter $\Delta
\left( r\right) $ and the effective charge $e^{*}$ in Eq. (4) equal to $2e$ 
\cite{13}. Within the framework of GL theory Bardeen and Stephen studied the
flux flow in a pinning free mixed state and obtained the flow resistance $%
\rho _f$ as \cite{14}

\begin{equation}
\frac{\rho _f}{\rho _n}=\frac{2\pi a^2B}{\Phi _0}=\left( \frac a\xi \right)
^2\frac B{B_{c2}}\approx \frac B{B_{c2}}\text{,}  \eqnum{5}
\end{equation}
where $a$ is the radius of normal core in the local model, nearly equal to
the G-L coherence length $\xi $ which can be expressed in terms of Gorkov's
derivation with BCS superconductivity as \cite{13}

\begin{equation}
\xi ^2\left( T\right) =\frac{\hslash ^2}{2m^{*}\left| \alpha \left( T\right)
\right| }=\frac{7\hslash ^2\zeta \left( 3\right) \epsilon _F\beta _c^2}{%
12\pi ^2m^{*}}\left( 1-\frac T{T_c}\right) ^{-1}\text{,}  \eqnum{6}
\end{equation}
with $\zeta \left( 3\right) $ the Rieman zeta function and $\beta _c\equiv
1/k_BT_c$.

For phonon mediated BCS superconductors the energy gap $\Delta \left(
T\right) $ near $T_c$ is shown as

\begin{equation}
\left( \frac{\Delta \left( T\right) }{\Delta _0}\right) ^2=3.1\left( 1-\frac %
T{T_c}\right) \text{,}  \eqnum{7}
\end{equation}
with $\Delta _0\equiv \Delta \left( T=0\right) \equiv 1.76k_BT_c$ \cite
{12,15}. Further study on the dependence of $\Delta $ on field and Cooper
pair velocity $v_s$ finds \cite{15}

\begin{eqnarray}
\left[ \beta _c\Delta \left( T,v_s\right) \right] ^2 &=&\frac{8\pi ^2}{%
7\zeta \left( 3\right) }\left( 1-\frac T{T_c}\right) -\frac 23\left( \beta
_c\hslash k_F\right) ^2v_s^2  \eqnum{8} \\
&=&\left[ \beta _c\Delta \left( T,0\right) \right] ^2-\frac 23\left( \beta
_c\hslash k_F\right) ^2v_s^2\text{.}  \nonumber
\end{eqnarray}

Inserting (6), (7), (8) into (5) and considering the identity

\begin{equation}
\ln \left( 1-x\right) =-\sum \frac{x^n}n,\text{ } 
\begin{array}{ll}
& 
\end{array}
n=1,2,...\text{, with}-1<x\leqslant 1\text{,}  \eqnum{9}
\end{equation}
We find

\begin{eqnarray}
\rho _f\left( T,B,v_s\right) &=&\rho _n\left( T,B\right) e^{-\gamma _1}\exp
\left[ \ln \frac{\Delta ^2\left( T,B,0\right) }{\Delta ^2\left(
T,B,v_s\right) }\right]  \nonumber \\
&\approx &\rho _n\left( T,B\right) \exp \left\{ -\gamma _1\left[ 1-\frac{%
2\left( \hslash k_F\right) ^2v_s^2}{3\gamma _0\Delta ^2\left( T,B,0\right) }%
\right] \right\} \text{.}  \eqnum{10}
\end{eqnarray}
Using the relation $J_s=en_sv_s$ with the density of superconducting
electrons $n_s\approx 2n\left( 1-T/T_c\right) \approx 2n\Delta ^2\left(
T\right) /\pi \Delta _0^2$, one gets Eq. (10) in the form

\begin{equation}
\rho _f\left( T,B,J\right) =\rho _n\left( T,B\right) \exp \left[ -\gamma
_1\left( 1-\frac{J_s^2}{J_d^2}\right) \right] \text{,}  \eqnum{11}
\end{equation}
with

\begin{equation}
J_d^2\approx \frac{2\gamma _0e^2n^2\Delta ^6\left( T,B,0\right) }{\left(
\hslash k_F\right) ^2\pi \Delta _0^4}\propto \gamma _1T_c^2\left( 1-\frac T{%
T_c}\right) ^3\text{.}  \eqnum{12}
\end{equation}
Eq. (11) is the same form of Eq. (2) in the pinning-free $\left( \gamma
_2=0\right) $ case with the evaluation of average supercurrent $J_s$ as

\begin{equation}
J_s=J-\frac{E\left( J\right) }{\rho _n\left( T,B\right) }\text{.}  \eqnum{13}
\end{equation}

III. RESISTANCE DUE TO FLUX CREEP

Working in the London limit, Nelson showed that according to the GL free
energy Eq. (4) a system of $N$ flux lines with a field $H$ along the $z$
direction in a sample length $L$ can be described with the free energy
represented by the trajectories $\left\{ \stackrel{\rightharpoonup }{r_j}%
\left( z\right) \right\} $ of these flux lines \cite{16,17,18}. Considering
further the pinning potential $V_P\left( \stackrel{\rightharpoonup }{r}%
\right) $ arising from inhomogeneities and defects in sample \cite{7,19,20},
the free energy of such a sample with $N$ flux lines can be expressed as

\begin{equation}
F=\frac 12\varepsilon _l\sum_{j=1}^N\int_0^L\left| \frac{d\stackrel{%
\rightharpoonup }{r_j}\left( z\right) }{dz}\right| ^2dz+\frac 12\sum_{i\neq
j}\int_0^LV\left( r_{ij}\right) dz+\sum_{j=1}^N\int_0^LV_P\left[ \stackrel{%
\rightharpoonup }{r_j}\left( z\right) \right] dz\text{,}  \eqnum{14}
\end{equation}
Here $V\left( r_{ij}\right) =V\left( \left| \stackrel{\rightharpoonup }{r_i}-%
\stackrel{\rightharpoonup }{r_j}\right| \right) =2\varepsilon _0K_0\left(
r_{ij}/\lambda _{ab}\right) $ is the interaction potential between lines
with in-plane London penetration depth $\lambda _{ab}$ and $K_0\left(
x\right) $ is the modified Bessel function $K_0\left( x\right) \approx
\left( \pi /2x\right) ^{1/2}e^{-x}$. $\varepsilon _l$ is the linear tension
of flux line and $\varepsilon _0\approx \left( \Phi _0/4\pi \lambda
_{ab}\right) ^2$ is the energy scale for the interaction.

Thermally activated flux motion is considered as the sequence of thermally
activated jumps of the vortex segments or vortex bundles between the
metastable states generated by disorder. Every elementary jump is viewed as
the nuclearation of the vortex loop, and the mean velocity of the vortex
system is determined by the nuclearation rate \cite{7,19,20}

\begin{equation}
v\varpropto \exp \left( \delta F/kT\right) \text{.}  \eqnum{15}
\end{equation}
Here $\delta F$ is the free energy for the formation of the critical size
loop or nucleus which can be found by means of the standard variational
procedure from the free energy functional due to the in-plane displacement $%
\stackrel{\rightharpoonup }{u}\left( z\right) $ of the moving vortex during
loop formation

\begin{equation}
F_{loop}\left[ \stackrel{\rightharpoonup }{u}\right] =\int dz\left[ \frac 12%
\varepsilon _l\left| \frac{d\stackrel{\rightharpoonup }{u}\left( z\right) }{%
dz}\right| ^2+V_P\left[ \stackrel{\rightharpoonup }{u}\left( z\right)
\right] -\left( f_L+f_\eta \right) \cdot \stackrel{\rightharpoonup }{u}%
\right] \text{,}  \eqnum{16}
\end{equation}
where $f_L=\stackrel{\rightharpoonup }{J}\times \stackrel{\rightharpoonup }{%
e_z}/c$ is the Lorentz force due to applied current $J$ and $f_\eta $ is the
viscous drag force on vortex, $f_\eta =-\eta v_{vortex}$, with $v_{vortex}=d%
\stackrel{\rightharpoonup }{u}/dt$ and viscous drag coefficient $\eta
\approx \left( \Phi _0B_{c2}\right) /\left( \rho _nc^2\right) $ as estimated
by Bardeen and Stephen \cite{14}.

Equation (16) is similar to the basic equations of collective pinning model
and vortex-glass as well as Bose-glass models \cite{21} for considering the
vortex dynamics at low temperatures. The only difference is the omission of $%
f_\eta $ in the latter, which is quite resonable since the velocity of
vortex $v_{vortex}$ usually is very small in the case of low temperature
flux creep, where one finds $f_\eta \ll f_L$. However, we have to consider
the term $f_\eta $ in our equation (16) not only because the fact that
viscous drag force does ever accompany the vortex displacement as shown
experimentally by Kunchur et al. \cite{22} but also for the need in our task
to describe the current-voltage characteristic of vortex system more
accurately for the case across the irreversibility line where $f_L$ and $%
f_\eta $ are comparable in magnitude, so now we have

\begin{equation}
F_{loop}\left[ \stackrel{\rightharpoonup }{u}\right] =\int dz\left[ \frac 12%
\varepsilon _l\left| \frac{d\stackrel{\rightharpoonup }{u}\left( z\right) }{%
dz}\right| ^2+V_P\left[ \stackrel{\rightharpoonup }{u}\left( z\right)
\right] -f_s\cdot \stackrel{\rightharpoonup }{u}\right] \text{,}  \eqnum{16'}
\end{equation}
with

\begin{equation}
f_s=f_L+f_\eta =\frac{J_P\Phi _0}c\times \stackrel{\rightharpoonup }{e_z} 
\begin{array}{lll}
&  & 
\end{array}
\text{and } 
\begin{array}{lll}
&  & 
\end{array}
J_P=J-\frac E{\rho _f}\text{,}  \eqnum{17}
\end{equation}
where we used the relation $E=\stackrel{\rightharpoonup }{v}\times \stackrel{%
\rightharpoonup }{B}$ and Eq. (5).

The free energy functional $F_{loop}\left[ \stackrel{\rightharpoonup }{u}%
\right] $ contains two parts. The last term in the right hand side of Eq.
(16) represents the net energy gained from the applied current which is
proportional to $J_p$ in Eq. (17) and the rest are corresponding to the
increase of the elastic energy during the loop formation. As argued by
Fisher et al. \cite{7}, in a bulk superconductor vortex lines are extended
one-dimensional (1D) objects and the response of vortices to an applied
current can be described by vortex loop formation of area $S=L_{\perp }\cdot
L_z\thicksim L_{\perp }^\kappa $ and elastic energy increase $\thicksim
U_c\cdot L_{\perp }^\theta $, where $L_{\perp }$ is the transverse
displacement of vortex-line segment of length $L_z$, with scaling relation $%
L_{\perp }\thicksim L_z^\zeta $.

Consequently, the free energy of loop formation can be estimated in the
dependence of $L_{\perp }$ as

\begin{equation}
F_{loop}\left( L_{\perp }\right) \thickapprox U_cL_{\perp }^\theta -J_p\frac{%
\Phi _0}cL_{\perp }^\kappa \text{.}  \eqnum{18}
\end{equation}
For a given value of $J_p$, $F_{loop}\left( L_{\perp }\right) $ first
increases and then decreases with increasing $L_{\perp }$ (since $\kappa
>\theta $ and $J_p<J_c$). The barrier energy is the maximum value of loop
formation free energy

\begin{equation}
\delta F=U\left( J_p\right) =F_{loop}\left[ L_{\perp }^{*}\left( J_p\right)
\right] \text{,}  \eqnum{19}
\end{equation}
with definition of $L_{\perp }^{*}\left( J_p\right) $

\begin{equation}
\left. \frac{\partial F_{loop}\left( L_{\perp }\right) }{\partial L_{\perp }}%
\right| _{L_{\perp }=L^{*}}=0\text{, } 
\begin{array}{lll}
&  & 
\end{array}
L_{\perp }^{*}\left( J_p\right) =\left( \frac{c\theta U_c}{\Phi _0\kappa J_p}%
\right) ^{1/\left( \kappa -\theta \right) }\text{.}  \eqnum{20}
\end{equation}
Combining Eqs. (18), (19) and (20) one finds

\begin{equation}
U\left( J_p\right) \thickapprox U_c\left( \frac{J_c}{J_p}\right) ^\mu \text{,%
}  \eqnum{21}
\end{equation}
where

\begin{equation}
J_c\thicksim \frac{\theta U_cc}{\kappa \Phi _0}\text{ } 
\begin{array}{lll}
&  & 
\end{array}
\text{and } 
\begin{array}{lll}
&  & 
\end{array}
\mu \thicksim \frac \theta {\kappa -\theta }=\frac{2\zeta -1}{2-\zeta }\text{%
.}  \eqnum{22}
\end{equation}
The current dependence (21) of the barrier $U$ to current-driven flux motion
implies a current-voltage characteristic of the form

\begin{equation}
E\left( J\right) =\rho _fJ\exp \left[ -\frac{U_c}{kT}\left( \frac{J_c}{J_p}%
\right) ^\mu \right] \text{.}  \eqnum{23}
\end{equation}
This form in one hand turns to the pinning-free flux flow resistive regime
as $J,J_p\gg J_c$ and clearly goes to zero in the other hand as $J_p$ $%
\thicksim J\rightarrow 0$, with no linear resistance term.

In real samples, the critical size of loop formation $L_{\perp }^{*}\left(
J_p\right) $, and thus $U\left( J_p\right) $ may always manifest a nonzero
resistance at sufficiently small measuring current with high enough accuracy
of measurements. Considering this real size effect one finds a general
normalized form of the current-voltage characteristic in the form (see
Appendix)

\begin{equation}
y=x\exp \left[ -\gamma \left( 1+y-x\right) ^p\right] \text{,}  \eqnum{24}
\end{equation}
with

\begin{equation}
\gamma =\frac{U_c}{kT}\left( \frac{J_c}{J_L}\right) \text{, } 
\begin{array}{lll}
&  & 
\end{array}
x=\frac J{J_L}\text{, } 
\begin{array}{lll}
&  & 
\end{array}
y=\frac{E\left( J\right) }{\rho _fJ_L}\text{, } 
\begin{array}{lll}
&  & 
\end{array}
p=\mu \text{.}  \eqnum{25}
\end{equation}
Where $J_L$ is the transport current density corresponding to the case where
the critical size of loop formation is equal to the sample size $L$ as

\begin{equation}
\left( \frac{c\theta U_c}{\Phi _0\kappa }\right) \left[ J_L-\frac{E\left(
J_L\right) }{\rho _f}\right] ^{-1}=J_c\left[ J_L-\frac{E\left( J_L\right) }{%
\rho _f}\right] ^{-1}\thickapprox \frac{J_c}{J_L}=L^{\kappa -\theta }\text{.}
\eqnum{26}
\end{equation}

Together with the former equation (11) in section II, equation (24) is just
the normalized form of the resistive transition equation (2) in the case of
temperatures below the melting point of vortex solid $T<T_2=T_m\left(
B\right) $ with $p_2=\mu $ and $J_{c0}\left( T,B\right) =J_L\left(
T,B\right) $.

IV. COMPARISON WITH EXPERIMENTS

In this section we compare the above derived resistive transition equation
of mixed state in the form of Eq. (2) with the experimental data of several
typical superconductors with high $T_c$.

A. $MgB_2$

The recently discovered superconducting $MgB_2$ has a high critical
temperature $T_c\thicksim 40K$ comparable to the cuprate $LaSrCuO$.
Meanwhile, it shows isotope effect of phonon-mediated superconductivity.
Finnemore et al. measured the transport and magnetic properties of sintered
pellet of $MgB_2$ and find the Ginzburg-Landau parameter $\kappa \thicksim
26 $ \cite{9}. In Fig. 1 we show the comparison of the resistive transition
equation (2) with their experimental data of the temperature dependent
resistance of $MgB_2$ from 300K to 1.9K in different applied fields.

B. $YBa_2Cu_3O_{7-\delta }$ (YBCO)

It is widely believed that the mechanism of superconductivity in high-$T_c$
cuprates may be essentially different, from the familiar s-wave type pairing
on which conventional BCS theory is based \cite{22}. Kwok et al. studied the
width and shape of the resistive transition of untwinned and twinned single
crystals of YBCO in fields up to $8T$ \cite{8} .This is one of the first
evidences which came from transport measurements for vortex melting in YBCO.
The samples are of high quality and near optimum doping confirmed by their
zero-field resistive transition of $R_{zero}>92.0K$ and $\Delta T_c\left(
10\%-90\%\right) <0.2K$. A ''knee'' in $R\left( T\right) $ curve is clearly
seen. We compare our resistive equation (2) with their experimental data of
untwinned YBCO crystal in Fig. 2.

Agreement is fair in both cases.

V. DISCUSSION

Though the resistive transition equation (2) fits the experimental data of
both $MgB_2$ and YBCO crystal, there are still essential differences in the
parameters entered into Eq. (2) for these two kinds of materials. In Fig. 1
we see the parameters used to fit the experimental data of $MgB_2$ by
Finnemore et al. \cite{8} follow the relations $I_d\propto \gamma
_1^{0.5}\left[ T_c\left( B\right) -T\right] ^{1.5}$ and $\gamma _1\equiv \ln
\left[ R_n\left( T,B\right) /R_f\left( T,B,J\rightarrow 0\right) \right]
=\ln \left[ B_{c2}\left( T\right) /B\right] $ in agreement with Eq. (17) and
the flux-flow resistance equation derived by Bardeen and Stephen \cite{4}
for BCS superconductivity. In contrast, the parameters to fit the
experimental data of YBCO crystal by Kwok et al. \cite{8}show the relations $%
I_d\propto \gamma _1^{0.5}\left[ 1-T/T_c\left( B\right) \right] ^{2.3}$ and $%
\gamma _1\equiv \ln \left[ R_n\left( T,B\right) /R_f\left( T,B,J\rightarrow
0\right) \right] =\ln \left[ B_{c2}\left( T\right) /B\right] ^m$, with $%
m=0.84B^{-0.59}-0.07$ which are somewhat different from Eq. (17) and the
results of Bardeen and Stephen \cite{4} in the exponents. At present a
correct microscopic theory for the high-$T_c$ cuprates is still a
challenging problem in condensed matter physics. While it seems that the
established superconductivity of cuperates is of $d_{x^2-y^2}$ symmetry, the
next question is whether it can be described by a BCS- like theory suitably
modified to in clude a d-wave gap or it is of some non-BCS origins \cite{23}%
. On the one hand, there are recent reports on the evidences for strong
electron-phonon coupling in high-$T_c$ cuprates by the analysis of
photoelectron spectra \cite{24} and the unconventional isotope effect \cite
{25} as well as the identification of the bulk pairing symmetry in high-$T_c$
hole-doped cuprates as extended s-wave symmetry with eight-line nodes and as
anisotropic s-wave in electron-doped cuprates \cite{26} which seem favouring
the BCS-like machanism. On the other hand, the well established unusual
properties of both the normal and superconducting states of cuprates seem
related to the fact that the cuprates are doped Mott insulators and
stimulate some ideas based on new exitations such as spin-charge separation 
\cite{27,28}, stripes \cite{29}, new symmetry relating superconductivity and
magnetism \cite{30} or quantum critical points \cite{31} which suggest a
non-BCS state. The interplay between theory and experiment promises to be
mutually benificial, in the best traditions of physics research. In this
paper we have compared the derived equation (2) only with the data of
optimally doped YBCO single crystal \cite{8}. However, the unusual
properties of cuprates appear even more striking in the underdoped region. A
thorough study on the resistive transition of underdoped cuprates may
provide new important insights into the nature of cuprates.

VI. SUMMARY

Starting from the Ginzburg-Landau functional we derived a resistive
transition equation for the mixed state of superconductors. This equation in
its general form agrees with the experimental data of superconducting $MgB_2$
pellet and optially doped untwinned YBCO single crystal but the parameters
to fit the data of cuprate are somewhat different from the prediction based
on BCS theory.

ACKNOWLEDGMENTS

This work is supported by the Ministry of Science and Technology of China
(NKBRSF-G 1999064602) and the Chinese NSF.

APPENDIX

In this appendix we provide the derivation of Eq. (24). Substituting Eq.
(17) into the bracket on the right-hand side of Eq. (23) and taking its
logarithm we get

\begin{equation}
J-J_f=\left( \frac{U_c}{kT}\right) ^{1/\mu }J_c\left[ \ln \left( \frac J{J_f}%
\right) \right] ^{-1/\mu }\text{,}  \eqnum{A.1}
\end{equation}
with $J_f\equiv E\left( J\right) /\rho _f$. Since the critical size of loop
formation $L_{\perp }^{*}$ is limited by the sample size $L$ as $L_{\perp
}^{*}\leqslant L$, one finds always

\begin{equation}
U\left( J\right) \leqslant U\left( J_L\right) \text{,}  \eqnum{A.2}
\end{equation}
with the definition

\begin{equation}
\left( \frac{c\theta U_c}{\Phi _0\kappa }\right) \left[ J_L-\frac{E\left(
J_L\right) }{\rho _f}\right] ^{-1}\equiv L^{\kappa -\theta }\text{.} 
\eqnum{A.3}
\end{equation}
Here we have used the relations in Eq. (17) and Eq. (20). Thus, Eq. (A.1)
can be expressed in the form

\begin{equation}
J-J_f=\left( \frac{U_c}{kT}\right) ^{1/\mu }J_c\left( 1+h\right) ^{-1}\left[
\ln \left( \frac{J_L}{J_{Lf}}\right) \right] ^{-1/\mu }\text{,}  \eqnum{A.4}
\end{equation}
where $J_{Lf}$ $\equiv E\left( J_L\right) /\rho _f$, which is much smaller
than $J_L$ and

\begin{equation}
h\equiv \frac{-\left[ \ln \left( J_L/J_{Lf}\right) ^{1/\mu }-\ln \left(
J/J_f\right) ^{1/\mu }\right] }{\ln \left( J_L/J_{Lf}\right) ^{1/\mu }}\text{%
, } 
\begin{array}{lll}
&  & 
\end{array}
\left| h\right| <1\text{.}  \eqnum{A.5}
\end{equation}
Using the approximation $\left( 1+h\right) ^{-1}\thickapprox 1-h$ for $%
\left| h\right| <1$, finally we find Eq. (A.1) in the form

\begin{equation}
x-y=1-\ln \left( \frac xy\right) ^{1/p}\gamma ^{-1/p}\text{,}  \eqnum{A.6}
\end{equation}
which is exactly the current-voltage characteristic Eq. (24) with the
definitions

\begin{eqnarray}
\gamma &\equiv &\ln \frac{J_L}{J_{Lf}}=\left( \frac{U_c}{kT}\right) \left( 
\frac{J_c}{J_L-J_{Lf}}\right) \thickapprox \left( \frac{U_c}{kT}\right)
\left( \frac{J_c}{J_L}\right) ^p\text{,}  \nonumber \\
x &\equiv &\left( \frac{U_c}{kT}\right) ^{-1/p}\left( \ln \frac{J_L}{J_{Lf}}%
\right) ^{1/p}\left( \frac J{J_c}\right) =\frac 12\frac J{J_L-J_{Lf}}%
\thickapprox \frac J{J_L}\text{,}  \nonumber \\
y &\equiv &\left( \frac{U_c}{kT}\right) ^{-1/p}\left( \ln \frac{J_L}{J_{Lf}}%
\right) ^{1/p}\left( \frac{E\left( J\right) }{J_0\rho _f}\right) =\frac 12%
\frac{E\left( J\right) }{\left( J_L-J_{Lf}\right) \rho _f}\thickapprox \frac{%
E\left( J\right) }{J_L\rho _f}\text{.}  \eqnum{A.7}
\end{eqnarray}

\begin{figure}[tbp]
\caption{Comparison of Eq.(19) with the experimental resistance data of MgB$%
_2$ samples measured by Finnemore et al. \protect\cite{8}. (a) A full view
in the applied field of $9T$. (b)Resistive transitions in different applied
fields. $\square \bigcirc \triangle \triangledown *\vartriangleleft
\vartriangleright +\bigstar \times $ denote experimental data and lines
denote the theoretical curves of Eq.(19) with corresponding applied fields.
The parameters in Eq.(19): 1 $U\left( T,B\right) \propto \left[ T_m\left(
B\right) -T\right] ^{0.8}\left( B+5.46\right) ^{-4.3}$. 2 $I_{c0}\propto
T_m\left( B\right) -T$; $I_d\propto \left[ T_c\left( B\right) -T\right]
^{1.5}\gamma _0^{0.5}$. 3 $T_m\left( B\right) =T_c\left( 0\right) \left[
1-\left( B/21.7\right) ^{0.84}\right] $, $T_c\left( B\right) =T_c\left(
0\right) \left( 1-B/22.4\right) $, $T_c\left( 0\right) =40.2K$. 4 $\gamma
_0\equiv \ln \left[ R_n\left( T,B\right) /R_f\left( T,B,J\rightarrow
0\right) \right] =\ln \left[ B_{c2}\left( T\right) /B\right] $, where $%
B_{c2}\left( T\right) =0.6\left[ T_c\left( 0\right) -T\right] ^{0.98}$. }
\label{Fig-1}
\end{figure}

\begin{figure}[tbp]
\caption{Comparison of Eq.(19) with the experimental resistive transition
data of untwinned YBCO crystal measured in different applied fields for $%
H_{//c}$ by Kwok et al.\protect\cite{8}. $\square \bigcirc \triangle
\triangledown *\vartriangleleft \vartriangleright +\bigstar \times $ denote
experimental data and lines denote the theoretical curves of Eq.(19) with
corresponding applied fields. The parameters in Eq.(19): 1 $U\left(
T,B\right) \propto \left[ T_m\left( B\right) -T\right] ^3B^{-4.22}$. 2 $%
I_{c0}\propto T_m\left( B\right) -T$; $I_d\propto T_c\left( B\right) \left[
1-T/T_c\left( B\right) \right] ^{2.3}\gamma _0^{0.5}$. 3 $T_m\left( B\right)
=T_c\left( 0\right) \left[ 1-\left( B/1200\right) ^{0.36}\right] $, $%
T_c\left( B\right) =T_c\left( 0\right) \left[ 1-\left( B/3.6\times
10^6\right) ^{0.3}\right] $, $T_c\left( 0\right) =97K$. 4 $\gamma _0\equiv
\ln \left[ R_n\left( T,B\right) /R_f\left( T,B,J\rightarrow 0\right) \right]
=\ln \left[ B_{c2}\left( T\right) /B\right] ^m$, where $m=0.84B^{-0.59}-0.07$
and $B_{c2}\left( T\right) =\left[ T_c\left( 0\right) -T\right] ^{3.3}$. }
\label{Fig-2}
\end{figure}

\end{document}